\documentclass[fleqn,twoside]{article}
\usepackage{espcrc2}

\usepackage{graphicx}
\usepackage[figuresright]{rotating}

\newcommand{\AmS}{{\protect\the\textfont2
  A\kern-.1667em\lower.5ex\hbox{M}\kern-.125emS}}

\title{Hybrid Activities of the Pierre Auger Observatory}

\author{Miguel Mostaf\'{a}\address{Physics Department,
  University of Utah\\Salt Lake City, UT 84112, U.S.A.}
        for the Pierre Auger Collaboration}
       
\begin{document}

\begin{abstract}
	The Pierre Auger Observatory detects ultra-high energy cosmic rays by implementing two complementary air-shower techniques. 
	The combination of a large ground array and fluorescence detectors, known as the \textit{hybrid} concept, 
means that a rich variety of measurements can be made on a single shower, 
providing much improved information over what is possible with either detector alone. 
In this paper I describe the hybrid reconstruction approach and the latest hybrid measurements.

\vspace{1pc}
\end{abstract}

\maketitle

\section{Introduction}

The energy spectrum of primary cosmic rays
extends over more than $12$ decades in energy.
It follows, over a large range, a simple power law
indicating its non-thermal nature.
The highest energy cosmic rays (which means particles with energies above $10$~EeV $\equiv 10^{19}$~eV)
must be produced by the most extreme non-thermal process in the Universe.
Their energy spectrum, arrival directions, and composition can be inferred from air shower observations.
However, the origin of these particles is still unknown and there is not a consensus whether there is 
a steepening in the energy spectrum at around $60$~EeV. 
Determining the shape of the spectrum in this energy range in sufficient detail 
requires an enormous detector exposure.
The Pierre Auger Observatory was designed to study these ultra-high energy cosmic rays.
Although under construction, the Auger Southern site is the largest cosmic ray detector in the world.

A fundamental characteristics of the Pierre Auger Observatory is its capability of \textit{hybrid} reconstruction of cosmic ray 
showers~\cite{Mostafa:2005kd}. 
Two independent detectors, the Surface Detector (SD), which samples the shower particles at ground, and the Fluorescence Detector (FD), 
which collects the fluorescence light emitted by the shower particles along their path in the atmosphere, 
are able to measure the energy and direction of the same cosmic ray shower.
The enhanced capabilities of the Auger hybrid detector are examined in this paper.

The Pierre Auger Observatory
was designed to observe, in coincidence, the shower particles at ground
and the associated fluorescence light generated in the atmosphere.
This is
achieved with a large array of water Cherenkov detectors coupled with
air-fluorescence detectors that overlook the surface array.
It is not simply a dual experiment.
Apart from important cross-checks and measurement redundancy,
the two techniques see air showers in complementary ways.
A single air shower is detected 3-dimensionally.
The ground array measures the 2-dimensional lateral structure of the shower at ground level,
with some ability to separate the electromagnetic and muon components.
The fluorescence detector records the longitudinal pro{f}ile of the shower during
its development through the atmosphere.

A \textit{hybrid} event is an air shower that is simultaneously detected
by the fluorescence detector and the ground array.
The Observatory was originally designed and is currently being built with a \textit{cross--triggering}
capability.
Data are recovered from both detectors whenever either system is triggered.
If an air shower independently triggers both detectors the event is tagged
accordingly.
There are also cases where the fluorescence detector,
having a lower energy threshold,
promotes a sub--threshold array trigger.
Surface stations are then matched by timing and location.
This is an important capability because these sub--threshold hybrid events
would not have triggered the array otherwise.
The geometrical reconstruction of the air shower's axis is accomplished by minimizing a
$\chi^{2}$ function involving data from all triggered elements in the eye and on the ground.
The reconstruction accuracy is far better than the ground array counters and the single eye
could achieve independently~\cite{Bonifazi:2005ns}.

The combination of the air fluorescence measurement and the particle detection on the ground provides
an absolute energy calibration.
The FD measurements determine the longitudinal development profile,
whose integral is proportional to the total energy of the electromagnetic particle cascade.
The SD independently estimates the shower energy by evaluating the particle density at $1000$~m from the core.
This ground parameter, called $S(1000)$, is estimated from the surface stations by fitting the measured densities to a lateral
distribution function~\cite{Ghia:2005bn}. 
A conversion factor that relates $S(1000)$ to the shower energy (based on the FD information) is extracted from hybrid events.
This reduces significantly the dependence on air shower models and on assumptions of the primary composition.

The Pierre Auger Collaboration started the construction of the Southern site in $2002$ in the city of 
Malarg\"{u}e, which is located at an elevation of $1400$~m a.s.l. in the province of Mendoza, Argentina.
After a successful operation of a prototype experiment~\cite{Abraham:2004dt},
the Southern Observatory started operation in \textit{hybrid} production mode in January, 2004.
Surface stations have a $100$\% duty cycle,
while fluorescence eyes can only operate on clear moonless nights.
Both surface and fluorescence detectors have been running simultaneously
$14\%$ of the time.
The number of hybrid events represents $10\%$ the statistics of the surface array data.
The Southern site will be completed in 2007.
Another important objective is to obtain a uniform exposure over the full sky.
This will be achieved by constructing a second instrument in the Northern hemisphere.
Each site is conceived to cover an area of $3000$~km$^{2}$ in order to collect about $1$ event
per week and site above $60$~EeV.
The Northern Observatory is planned to be sited in the U.S.

\section{Hybrid Geometrical Reconstruction}

A hybrid detector achieves the best geometrical accuracy by using timing information from all the detector components,
both FD pixels and SD counters.
Each element records a pulse of light from which one can determine the central time of the pulse and its uncertainty.
Each trial geometry for the shower axis yields a prediction for the times at each detector component.
Differences between actual and predicted times are weighted using their corresponding uncertainties,
squared, and summed to construct a $\chi^{2}$ value.
The hypothesis with the minimum value of $\chi^{2}$ is the reconstructed shower axis.

\begin{figure}[!ht]
\includegraphics[width=0.45\textwidth]{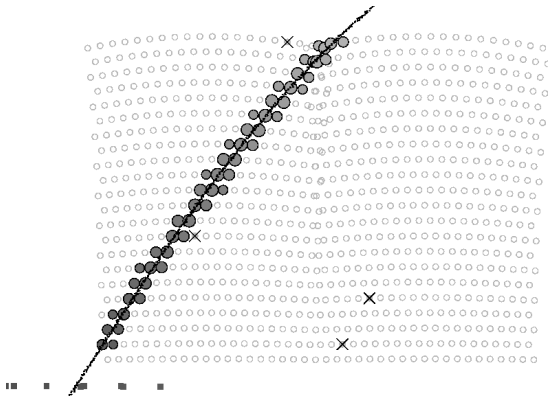}
\caption{
Light track of a hybrid event as seen by two adjacent fluorescence telescopes.
%The different colors indicate the timing sequence of the triggered pixels.
The full line is the fitted shower-detector plane.
The squares represent the surface detectors that also triggered in this event.%
}
\label{fig:Gcamera} 
\end{figure}

\begin{figure}[!ht]
\includegraphics[width=0.45\textwidth]{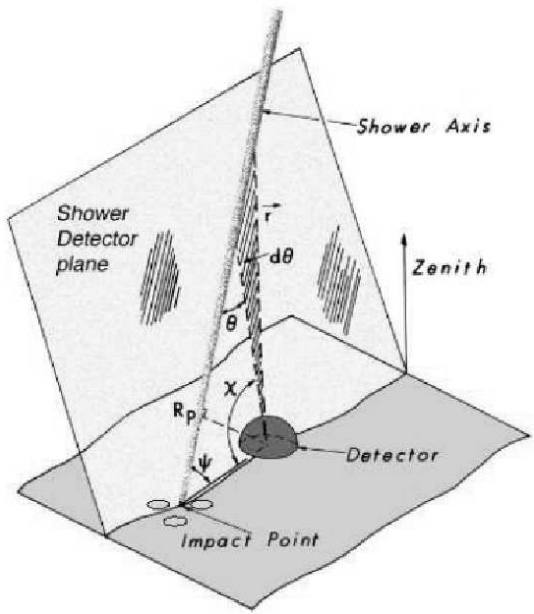}
\caption{Illustration of the geometrical shower reconstruction from the observables of the fluorescence detector.
}
\label{fig:sketch} 
\end{figure}

In the FD, cosmic ray showers are detected as a sequence of triggered pixels in the camera.
An example of an event propagating through two adjacent FD telescopes is presented in Fig.~\ref{fig:Gcamera}.
The first step in the analysis is the determination of the shower-detector plane (SDP).
The SDP is the plane that includes the location of the eye and the line of the shower axis.
(See the sketch in Fig.~\ref{fig:sketch}.)
Experimentally, it is the plane through the eye which most nearly contains the pointing directions of the FD pixels
centered on the shower axis.
(See fitted line in Fig.~\ref{fig:Gcamera}.)
Using a known axis provided from the Central Laser Facility (CLF), described in Ref.~\cite{Arqueros:2005yn},
the SDP reconstruction error can be evaluated by comparing the space angle between
the normal vector to the experimentally determined SDP and the known true normal vector.
This uncertainty in the SDP is not greater than about $0.1^{\circ}$.

Next, the timing information of the pixels is used for reconstructing the shower axis within the SDP.
As illustrated in Fig.~\ref{fig:sketch}, the shower axis can be characterized by two parameters:
the perpendicular distance $R_{p}$ from the eye to the track,
and the angle $\psi$ that the track makes with the horizontal line in the SDP.
Each pixel which observes the track has a pointing direction which makes an angle
$\chi_{i}$ with the horizontal line.
Let $t_{0}$ be the time when the shower front on the axis passes the point of closest approach $R_{p}$ to the eye.
The light arrives at the $i$th pixel at the time 
\begin{equation}
	t_i = t_0 + \frac{R_p}{c} \cot{[(\psi+\chi_i)/2]}.
	\label{eq:timefit}
\end{equation}

\begin{figure}[!ht]
\includegraphics[width=0.45\textwidth]{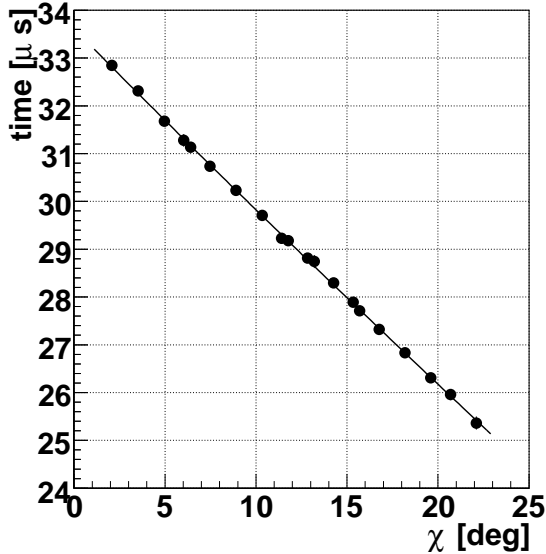}
\caption{Functional form that correlates the time of arrival of the light at each pixel with the 
angle between the pointing direction of that particular pixel and the horizontal line within the shower-detector plane.
}
\label{fig:timefit} 
\end{figure}

\begin{figure}[!ht]
\includegraphics[width=0.45\textwidth]{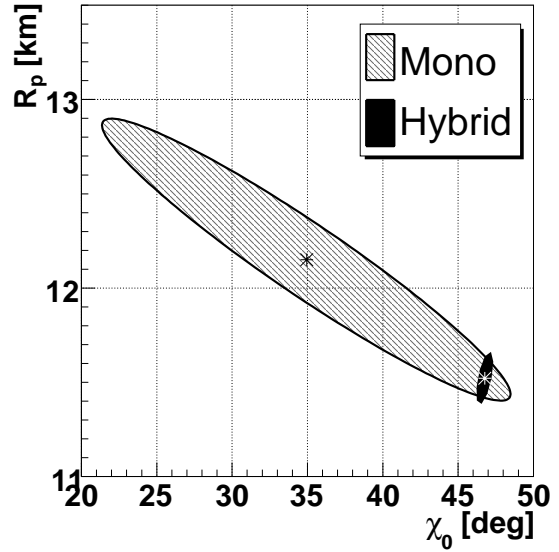}
\caption{Solutions (and $1\sigma$ regions) for the axis fit shown in Fig.~\ref{fig:timefit}.
The large uncertainty (and strong correlation) of the monocular reconstruction is broken using the 
timing information from the surface detectors (Hybrid).
The stars mark the solutions that minimize the $\chi^2$ for the axis reconstruction.
(Following the notation of this paper,
$\chi_0 = 180^{\circ} - \psi$.)
}
\label{fig:correlation} 
\end{figure}

The shower parameters are then determined by fitting the data points to this functional form.
Using the FADC electronics, such a monocular reconstruction may achieve excellent accuracy.
However, the accuracy of the monocular reconstruction is limited when the measured angular speed
$d\chi/dt$ does not change much over the observed track length.
An example is shown in Fig.~\ref{fig:timefit}.
For these events (usually short tracks) there is a small curvature in the functional form of Eq.~(\ref{eq:timefit})
such that there is a family of possible $(R_p,\psi)$ axis solutions.
This is also illustrated in Fig.~\ref{fig:correlation}.
Not only the uncertainties in $R_p$ and $\psi$ are large,
but also they are strongly correlated.
This translates directly into an uncertainty in the other shower parameters,
especially in the reconstructed shower energy.
This asymmetric uncertainty in the energy and angular resolution are important drawbacks of the
monocular reconstruction.

The aforementioned degeneracy can be broken by combining
the timing information from the SD stations with that of the FD telescopes.
This is called the \textit{hybrid} reconstruction.
The hybrid solution 
for the example shown in Fig.~\ref{fig:timefit} 
is shown in Fig.~\ref{fig:correlation}
as a white star and its uncertainty as the smaller (full) ellipse.

Since the SD operates with a $100\%$ duty cycle,
most of the events observed by the FD are in fact hybrid events.
There are also cases where the fluorescence detector,
having a lower energy threshold,
promotes a sub--threshold array trigger.
Surface stations are then matched by timing and location.
This is an important capability because these sub--threshold hybrid events
would not have triggered the array otherwise.
In fact, the time of arrival at a single counter at ground can suffice for the
hybrid reconstruction. 

\begin{figure}[!ht]
\includegraphics[width=0.45\textwidth]{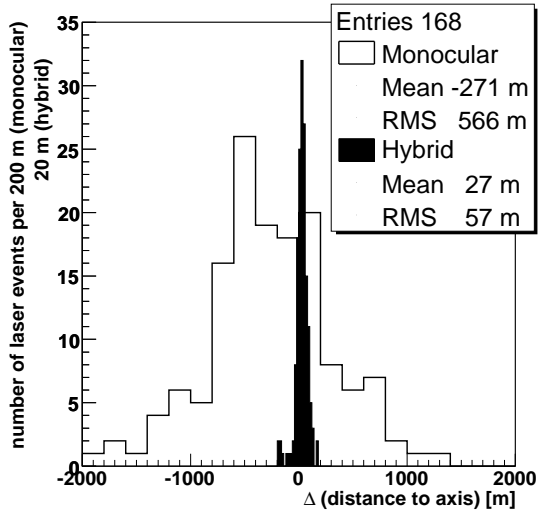}
\caption{\label{RESOLVRP} Difference between the reconstructed and true distance
from the eye to the vertical laser beam using the monocular and hybrid techniques.
The location of the laser is known to $5$~m.
}
\end{figure}

\begin{figure}[!ht]
\includegraphics[width=0.45\textwidth]{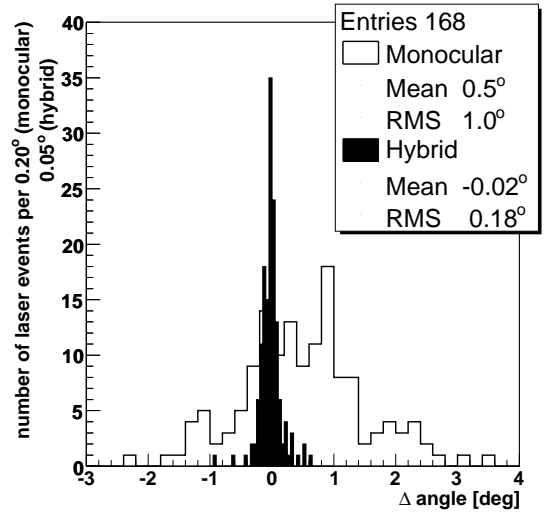}
\caption{\label{RESOLVCHI0} Angular difference between reconstructed and true direction
of the laser beam using the monocular and hybrid techniques.
The laser beam is vertical within $0.01^{\circ}$.
}
\end{figure}

The reconstruction uncertainties are evaluated
using events with \textit{known} geometries, \textit{i.e.} laser beams.
Since the location of the CLF 
(approximately equidistant from the first three fluorescence sites)
and the direction of the laser beam are
known to an accuracy better than the expected angular resolution of the
fluorescence detector, laser shots from the CLF can be used to measure the accuracy of
the geometrical reconstruction.
Furthermore, the laser beam is split and part of the laser light is sent
through an optical fiber to a nearby ground array station.
Thus, the axis of the laser light can be reconstructed both in monocular mode and in the
\textit{single-tank} hybrid mode.
The resolution of the monocular and hybrid reconstructions are compared in
Fig.~\ref{RESOLVRP} for the distance between the eye and the CLF, and
in Fig.~\ref{RESOLVCHI0} for the angle of the axis.
The results are very encouraging.
With the monocular reconstruction, the location of the CLF can be determined
with a resolution of $\sim500$~m.
After including the timing information of the single water tank,
the resolution improves by one order of magnitude with no systematic shift.

\begin{figure}[!ht]
\includegraphics[width=0.45\textwidth]{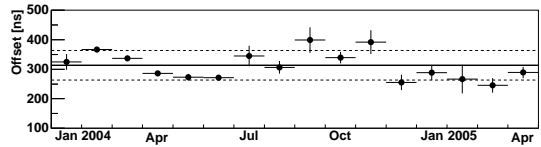}
\caption{\label{OFFSET} Time offset between the surface detector and one of the fluorescence detectors.
The variation in time is within the uncertainty of $50$~ns.
}
\end{figure}

As mentioned before, the laser light from the CLF produces simultaneous triggers in both the surface and
(three) fluorescence detectors. 
The recorded event times are used to measure and monitor the
relative timing between the two detectors.
The time offset between the first fluorescence eye and the surface detector is shown in
Fig.~\ref{OFFSET}.
This time offset is crucial for the accuracy of the hybrid reconstruction, 
and it has been measured to
better than $50$~ns~\cite{Allison:2005vj}.
The contribution to the systematic uncertainty in the core location due
to the uncertainty in the time synchronization is $20$~m.

Using the timing information from the eye's pixels together with the surface stations
to reconstruct real air showers,
a core location resolution of $50$~m is achieved.
The resolution for the arrival direction of cosmic rays is
$0.6^{\circ}$~\cite{Bonifazi:2005ns}.
These results for the \textit{hybrid} accuracy are in good agreement with
estimations using analytic arguments~\cite{Sommers:1995dm},
measurements on real data using a bootstrap method~\cite{Fick:2003qp},
and previous simulation studies~\cite{Dawson:1996ci}.

\section{Energy Determination}

\begin{figure}[!ht]
\includegraphics[width=0.45\textwidth]{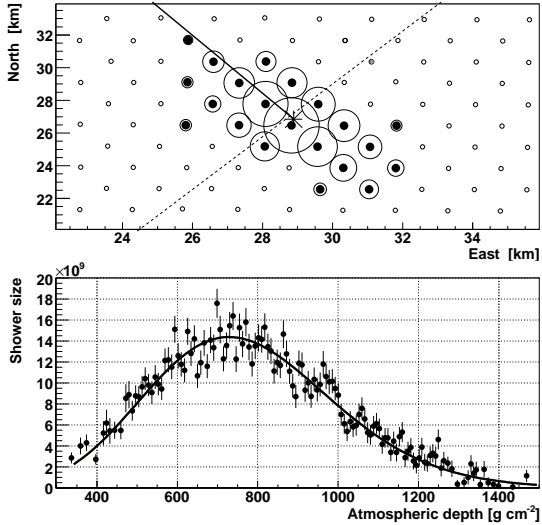}
\caption{\label{fig:profile} An air-shower event detected both in water tanks (\textit{top:} array view), 
and in a fluorescence detector (\textit{bottom:} longitudinal profile).
}
\end{figure}

After successful reconstruction of the event geometry,
the FADC traces of the FD pixels are analyzed in order to obtain the light emitted along the shower axis.
An atmospheric scattering model is used in this step to transform the light received at the FD back to the light emitted
from the shower axis.
The geometrical height, as observed by the telescopes, is converted to grammage of atmosphere.
The amount of fluorescence light emitted from a volume of air is proportional to the energy dissipated by the shower
particles in that volume.
The observed longitudinal light profile represents the energy loss in the atmosphere, which in turn is proportional
to the number of charged particles in a given volume.
The result for a hybrid shower of $\theta\sim60^{\circ}$ 
is shown in Fig.~\ref{fig:profile} (bottom).
The line represents the best fit to a Gaisser-Hillas function~\cite{GH},
yielding a primary energy of $E=(23\pm6)$~EeV, in good agreement with the $S(1000)$ determination. 

\section{Observatory Status}

\begin{figure}[!ht]
\centerline{\includegraphics[width=0.45\textwidth]{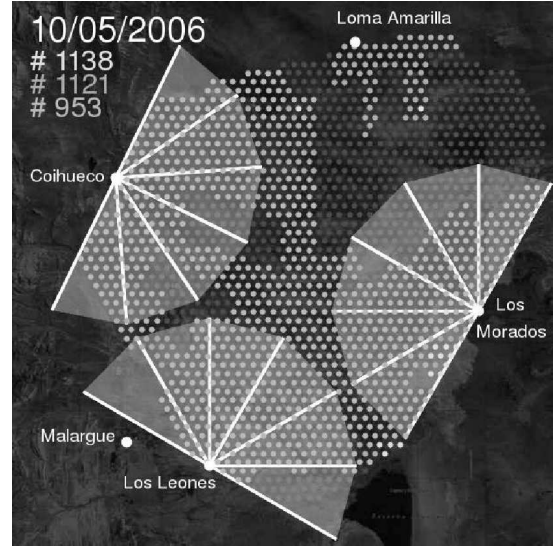}}
\caption{\label{STATUS} Status of the Pierre Auger Southern Observatory on May 10, 2006.
%Each color dot represents a surface station:
%\textit{red} means a proposed location,
%\textit{yellow} a deployed tank,
%\textit{orange} a tank with water (but without electronics), and
%\textit{green} a fully operational tank.
There were 1138 deployed tanks
and 953 of them were fully operational.
Also shown are the locations of 
the fluorescence detectors with their fields of view 
(for the three fully operational detectors).
The fourth fluorescence site (Loma Amarilla) is under construction.
}
\end{figure}

The status of the Southern Observatory on May 10, 2006 is summarized in Fig.~\ref{STATUS}.
There were 953 fully operational tanks at that time.
The {f}irst two fluorescence sites (Los Leones and Coihueco) were fully operational,
\textit{i.e.} running six telescopes each, in June, 2004.
The third site (Los Morados) started operation on March 18, 2005.
The fourth and last site (Loma Amarilla) is currently under construction and is scheduled
to start operation by the end of 2006.
The present average rate is $50$ hybrid events per night per eye,
for a total of $\sim40000$ events up to June, 2006.
At this rate, $4000$ hybrid events per month are expected
when the Southern site is completed.

\section{Hybrid Measurements}

In the Pierre Auger Observatory two complementary experimental approaches
are combined on a shower-by-shower basis within one single experiment.
Such redundancy allows cross-correlations between experimental techniques,
thereby controlling the systematic uncertainties.
Furthermore, there is an improvement in the resolution of the energy, mass,
and arrival direction of the reconstructed primary particles.
Data are being used to verify both the performance of the individual components~\cite{Bellido:2005aw,Bertou:2005bx},
as well as to produce the highest quality reconstructed air-shower events~\cite{Matthews:2005ve}.
Results are very promising~\cite{Mantsch:2005uq}
and underline the advantages of the hybrid approach.

\subsection{Energy Spectrum}

Hybrid air shower measurements 
are utilized in the spectrum analysis to avoid dependence on
specific numerical simulations of air showers and detector
responses to them.  The analysis is also free of assumptions
about the primary nuclear masses.  The FD
provides a nearly calorimetric, model-independent energy
measurement: fluorescence light is produced in proportion to
energy dissipation by a shower in the atmosphere.
Hybrid data establish the relation of shower energy to a ground parameter. 
%, which is the water Cherenkov signal in the SD at a distance of 1000 meters from the shower axis.  
Moreover,
hybrid data determine the trigger probability for individual
tanks as a function of core distance and energy, from which it
is found that the SD event trigger is fully efficient above $3$~EeV
for zenith angles less than $60^{\circ}$.  The SD exposure is
then calculated simply by integrating the geometric aperture over
time.
It is the continuously operating surface array which provides the high
statistics with unambiguous exposure.
Assigning energies to the SD event set is a two-step process.
The first step is to assign an energy parameter, $S_{38}$, to each
event.  Then the hybrid events are used to establish
the rule for converting $S_{38}$ to energy.

\begin{figure}[!ht]
\includegraphics*[width=0.45\textwidth]{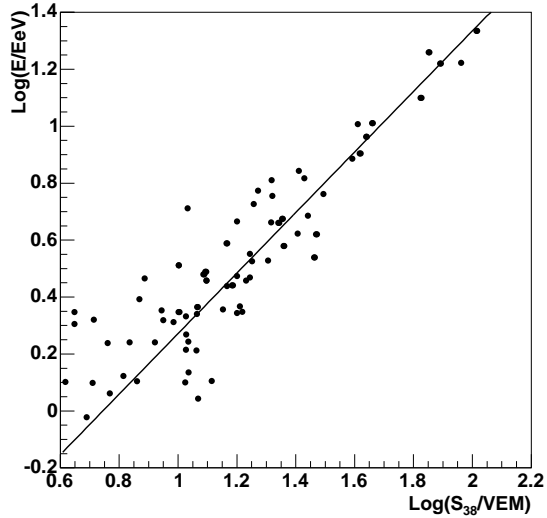}
\caption{\label{fig:cic} Shower energy (measured from the longitudinal profile) as a function of the ground parameter
  $S_{38}$.  Each point is a hybrid event recorded when
  there were contemporaneous aerosol measurements, whose 
  longitudinal profile included shower maximum in a measured
  range of at least 350~g\ cm$^{-2}$, and in which there was less than 10\%
  Cherenkov contamination.  The fitted line is
  $\log(E)=-0.79+1.06\ \log(S_{38})$.}
\end{figure}

The energy parameter $S_{38}$ for each shower comes from its
experimentally measured $S(1000)$.
%, which is the time-integrated water Cherenkov signal that would be measured by a tank 1000 meters from the core.  
%This ground parameter %is determined
%accurately by non-linear interpolation even when there is no tank
%at that particular core distance.
It 
may be regarded as the $S(1000)$
measurement the shower would have produced if it had arrived
$38^{\circ}$ from the zenith.
As it can be seen in Fig.~\ref{fig:cic}, 
$S_{38}$ is well correlated with the FD energy measurements of 
high quality hybrid events.
The fitted line gives an empirical rule for
assigning energies (in EeV) based on $S_{38}$ (in VEM):
\begin{equation}E = 0.16 \times S_{38}^{1.06}.\end{equation}
%The uncertainty in this rule is discussed below.  
The SD acceptance is not saturated below $3$~EeV. 
But the hybrid events used in Fig.~\ref{fig:cic},  
which start at $\sim1$~EeV,  
are those
with core locations and arrival directions such that they have
a probability greater than 0.9 for satisfying the SD trigger and
quality conditions.  
%These events increase the statistics and the
%``moment arm'' of the correlation without introducing appreciable
%bias.
%
The distribution over energy produced by this two-step procedure
becomes the cosmic ray spectrum presented in Ref.~\cite{Sommers:2005vs}.

\subsection{Anisotropy}

Due to the much improved angular accuracy,
the \textit{hybrid} data sample is ideal for anisotropy studies
and, in particular, for point source searches.
Results on a search for a point--like source in the direction of the Galactic
Center (GC) using these hybrid events were presented in Ref.~\cite{Aglietta:2006ur}.
These events have a better angular resolution %\cite{Bonifazi:2005ns}
($ 0.7^\circ$ at 68\% c.l. in the EeV energy range).
Considering the hybrid events with an energy between $10^{17.9}$~eV and $10^{18.5}$~eV,
no significant excess is seen in the GC direction.
For instance, in an optimal top-hat window of
$1.59\sigma\simeq 0.75^\circ$ radius,
 0.3 events are expected while
no single event direction falls within that circle.
This leads to a source flux upper-bound at 95\% c.l. of
\begin{equation}
\Phi_s^{95}=\xi\; 0.15 \ {\rm km^{-2}\ yr^{-1}}.
\end{equation}
%which is about a factor of two weaker than the SD flux bound.
(The factor $\xi$ is close to unity and parametrises the
uncertainties in the flux normalization.)

Note that
the energy assignments of the FD apply regardless of the assumed primary
composition (except for a small correction to account for the missing energy),
be they protons or heavy nuclei. However,
the acceptance has  a dependence on composition because different primaries
develop at different depths in the atmosphere. Since a quality
requirement for hybrid events is to have the maximum of the shower
development inside the field of view of the telescopes, this affects
the sensitivity to different primaries.
The bound obtained is indeed conservative even if the bulk of the cosmic rays are
heavy nuclei.

\subsection{Composition}

An upper limit for the photon fraction in cosmic rays with energies above $10^{19}$~eV
was derived using the hybrid data set~\cite{Abraham:2006ar}.
The importance of the photon fraction lies in the fact that \textit{top-down}
models predict a considerable proportion of photons among the generated particles.
The measured photon flux is thus a valuable indicator of these non-accelerator models.

The method used in Ref.~\cite{Abraham:2006ar} to distinguish between hadrons and photons
in the hybrid data exploits the information on the longitudinal profile of the air shower.
In fact, this is the first such limit on photons obtained by observing the fluorescence light
profile of air showers.
The atmospheric depth at the shower maximum, $X_{max}$, is commonly used as a discriminant observable
for the cosmic ray composition because
lighter nuclei penetrate, on average, more deeply into the atmosphere.
Above $10^{19}$~eV, showers initiated by photons develop significantly deeper in the atmosphere than hadronic
showers. %, yielding $X_{max}$ values well above those of proton showers.

No candidate for a primary photon was found in the hybrid data taken between Jan, 2004 and Feb, 2006.
By comparing the observed $X_{max}$ of each hybrid event to predictions from hadronic simulations,
an upper limit of $16\%$ (at $95\%$ c.l.) was derived.
This results confirms and improves the existing limits above $10$~EeV.

This analysis is currently limited mainly by the the small number of events.
The number of hybrid events will considerably increase in the next few years,
and much lower primary photon fractions can be tested.
An upper limit of $\sim5\%$ could be achieved, for example, with two more years of data taking.
Moreover, the larger statistics will allow to increase the threshold energy
above $10$~EeV where even larger photon fractions are predicted by the models.
A similar limit ($\sim15\%$) but at higher energy (e.g. above $40$~EeV), would be well below existing limits
and severely constrain non-acceleration models.

\section{Conclusions}

The Pierre Auger Observatory is a hybrid detector with excellent capabilities
for studying the highest energy cosmic rays.
Much of its capability stems from the accurate geometric reconstruction it achieves.
The shower geometry is reconstructed combining information from the eyes and the ground detectors.
Arrival directions are determined to a small fraction of a degree and the shower core is located to an
accuracy of about $60$~m~\cite{Bellido:2005aw,Bonifazi:2005ns}.
%The accuracy of the hybrid reconstruction 
%is better than the one the ground array counters
%or the single eye could achieve independently.

The construction of the Southern Observatory is well under way.
Eighteen FD telescopes and more than $60\%$ of the surface array are in operation
taking data routinely.
At the present rate of deployment,
construction will be finish in mid 2007.
Detectors are performing very well and the first results are very encouraging.

Emphasis is placed on hybrid analysis that provide unprecedented quality
in geometry, energy, and mass reconstruction.
Of utmost importance for the near future will be the determination of the energy spectrum
to study the GZK feature, and the search for anisotropies in arrival directions.
It is important to note that
both SD and FD techniques have different systematics,
and results are preliminary at this stage while the Observatory is under construction.
The possibility of studying
the same set of air showers with two independent methods is
valuable in understanding the strengths and limitations of each technique.
The hybrid analysis bene{f}its from the calorimetry of the fluorescence technique
and the uniformity of the surface detector aperture.

In parallel to the completion of the Southern Observatory and to the analysis of data towards
the first scientific results, R\&D has started for the development of the Northern site in the US.

\end{document}